\documentclass[12pt,a4paper]{article}
\usepackage{graphicx}
\usepackage[T1]{fontenc}
\usepackage[utf8]{inputenc}
\usepackage{textcomp}
\usepackage[sc,osf]{mathpazo}
\usepackage{a4wide}  
\usepackage{latexsym,amsthm,amsfonts,amsmath,mathrsfs,amssymb}
\usepackage{dsfont}
\usepackage{accents}
\usepackage[nosort]{cite}
\usepackage{booktabs} 
\usepackage[unicode,implicit]{hyperref}
\hypersetup{%
  pdftitle    = {Hairy black holes, scalar charges and extended thermodynamics}
  pdfkeywords = {Black holes, scalar fields, scalar charges, hair, no-hair theorems},
  pdfauthor   = {Romina Ballesteros and Tom\'as Ort\'{\i}n},
  plainpages  = true,
  colorlinks  = true,
  citecolor   = blue,
  urlcolor    = red,
  linkcolor   = black
}
\newcommand{\hepth}[1]{{\tt
\href{http://www.arXiv.org/abs/hep-th/#1}{hep-th/#1}}}
\newcommand{\grqc}[1]{{\tt
\href{http://www.arXiv.org/abs/gr-qc/#1}{gr-qc/#1}}}

\newcommand{\arxiv}[1]{{\tt arXiv:\href{http://www.arXiv.org/abs/#1}{#1}}}
\newcommand{\doi}[1]{{\tt DOI:\href{http://dx.doi.org/#1}{#1}}}

\allowdisplaybreaks

\makeatletter
\@addtoreset{equation}{section}
\makeatother

\begin{document}

\begin{flushright}
\small
IFT-UAM/CSIC-23-088\\
August 9\textsuperscript{th}, 2023\\
\normalsize
\end{flushright}

\vspace{1cm}

\begin{center}

  {\Large {\bf Hairy black holes, scalar charges and extended thermodynamics}}

\vspace{1.5cm}

\renewcommand{\thefootnote}{\alph{footnote}}

{\sl\large Romina Ballesteros$^{1,2,}$}\footnote{Email: {\tt romina.ballesteros[at]estudiante.uam.es}}
{\sl\large and Tom\'as Ort\'{\i}n}$^{1,}$\footnote{Email: {\tt  tomas.ortin[at]csic.es}}

\setcounter{footnote}{0}
\renewcommand{\thefootnote}{\arabic{footnote}}
\vspace{1cm}

${}^{1}${\it Instituto de F\'{\i}sica Te\'orica UAM/CSIC\\
C/ Nicol\'as Cabrera, 13--15,  C.U.~Cantoblanco, E-28049 Madrid, Spain}\\

\vspace{.5cm}

$^{2}${\it Pontificia Universidad Cat\'olica de Valpara\'{\i}so,
  Instituto de F\'{\i}sica, Av. Brasil 2950, Valpara\'{\i}so, Chile}

\vspace{1cm}


{\bf Abstract}
\end{center}
\begin{quotation}
  {\small We explore the use of the recently defined scalar charge which
    satisfies a Gauss law in stationary spacetimes, in the context of theories
    with a scalar potential. We find new conditions that this potential has to
    satisfy in order to allow for static, asymptotically-flat black-hole
    solutions with regular horizons and non-trivial scalar field. These
    conditions are equivalent to some of the known ``no-hair'' theorems (such
    as Bekenstein's). We study the extended thermodynamics of these systems,
    deriving a first law and a Smarr formula. As an example, we study the
    Anabal\'on-Oliva hairy black hole.}
\end{quotation}

\newpage
\pagestyle{plain}




\section{Introduction}

One of the most remarkable aspects of black holes is the fact that all their
properties are determined by their conserved charges, irrespectively of their
formation history. This fact is referred to as \textit{uniqueness} (there is
only one black hole for a given set of conserved charges) or absence of
\textit{hair} (there are no other parameters apart from the conserved charges
characterizing different black holes) and it can be argued that it lies at the
very heart of the thermodynamic interpretation of the dynamics of black holes.

This property has been tested in theories in which matter fields giving rise
to different conserved charges are coupled to gravity. We can distinguish two
broad types:

\begin{enumerate}

\item Matter fields with gauge symmetries, such as 1-forms in 4
  dimensions. The conserved charges associated to these symmetries (electric
  and magnetic charges) are defined through surface integrals and are believed
  to be preserved by quantum gravity.
  
\item Matter fields with global symmetries, such as scalar
  fields.\footnote{The symmetries may act on other fields as well.} The
  conserved charges associated to these symmetries are defined through volume
  integrals and are believed not to be preserved by quantum
  gravity. Therefore, black holes should not be characterized by this kind of
  charge, and possible non-trivial fields of this kind in black-hole
  spacetimes would, then, be understood as ``hair'' violating uniqueness.

\end{enumerate}

It goes without saying that it is the second type of matter and the possible
violations of black-hole uniqueness that it may induce that has attracted most
interest. It is also the subject of this work.

In order to discuss black holes with scalar hair we first have to characterize
scalar hair more precisely. The most naive way to do it would be to use the
conserved charge associated to the global symmetries of the theory that act on
the scalars. There are several reasons why this is not possible, even though
the contrary is sometimes assumed in the literature:

\begin{enumerate}

\item There may not be any global symmetry acting on the scalars at all and,
  therefore, there may not be an associated conserved charge. This is what
  actually happens in the theories considered in this paper in which the shift
  symmetry of a real scalar is broken by a scalar potential. In some works,
  the charge that would be conserved in absence of the potential is used, even
  though it is obviously not conserved and does not satisfy a Gauss law. The
  main problem with this kind of definitions comes from the next point,
  though.
 
\item In static black-hole spacetimes the volume integral that gives the
  globally conserved charge usually vanishes when integrated over a spacelike
  hypersurface \cite{Ballesteros:2023iqb}.
  
\end{enumerate}

For these reasons, in most of the literature it has been customary to use a
definition of scalar charge based on the asymptotic expansion of the scalar:
the scalar charge would be given, up to normalization, by the coefficient of
the $1/r$ term in that expansion (see, for instance,
Ref.~\cite{Gibbons:1996af}). This definition can be used in simple settings
but it is clear that a coordinate-dependent definition is necessary to study
the properties of this charge and establish general results.

In Refs.~\cite{Pacilio:2018gom,Ballesteros:2023iqb} a covariant definition of
scalar charge of a stationary black hole as the integral of an on-shell closed
2-form was proposed. In the cases considered so far, this definition gives the
same value as the conventional definition based on the asymptotic expansion,
but with the new definition one can go farther: the closedness of the 2-form
charge implies that this scalar charge satisfies a Gauss law\footnote{It is
  worth stressing that by no means this implies that this scalar charge is
  conserved.} and the covariant definition can be used to recover the scalar
term in the first law of black-hole mechanics found in
Ref.~\cite{Gibbons:1996af}.

One of the empirical properties of this scalar charge is that, in black-hole
spacetimes, it is usually completely determined by the conserved charges and
asymptotic values of the scalars. In the language of
Ref.~\cite{Coleman:1991ku} this kind of scalar charge corresponds to
``secondary hair'' and the black hole is still completely determined by the
values of its truly conserved charges (plus the asymptotic values of the
scalars). Whenever there are solutions with the same conserved charges but the
scalar charge does not have that value (a particular function of the conserved
charges and asymptotic values of the scalars) but is a free parameter that
describes ``primary hair'' in the language of Ref.~\cite{Coleman:1991ku}, the
solution does not have a regular horizon and does not describe a black hole.
This is illustrated by the solutions in
Refs.~\cite{Janis:1968zz,Agnese:1994zx}. The covariant definition of scalar
charge of Refs.~\cite{Pacilio:2018gom,Ballesteros:2023iqb} can be used to
determine the particular value of the scalar charge allowed in presence of a
bifurcate black-hole horizon, which is, as a matter of fact, equivalent to a
``no-hair theorem''.

In this paper we want to study the extensions of the results obtained in
Refs.~\cite{Pacilio:2018gom,Ballesteros:2023iqb} to the case in which a real
scalar is coupled to itself via a scalar potential instead of being coupled to
vector fields.  This is a very simple case which has been very much studied in
the past and several ``no-hair theorems'' have been proven for more or less
general classes of scalar potentials in
Refs.~\cite{Bekenstein:1972ny,Bekenstein:1995un,kn:HeuslerStraumann1992,Heusler:1992ss,Sudarsky:1995zg,Heusler:1996ft}.\footnote{See
  Ref.~\cite{Herdeiro:2015waa} for a review on the topic of hairy black holes
  with many references.}

One of the main assumptions in the proofs of these theorems is the positivity
of the scalar potential, related to the energy conditions and, not
surprisingly, asymptotically-flat black-hole solutions with scalar hair have
been found in theories whose scalar potential violates that condition
\cite{Bechmann:1995sa,Dennhardt:1996cz,Nucamendi:1995ex,Anabalon:2012ih,Cadoni:2015gfa}.
These solutions and their thermodynamics have not been studied from the point
of view of their scalar charges\footnote{In Ref.~\cite{Astefanesei:2023sep}
  the extended thermodynamics of some of these solutions has been studied, but
  not from the point of view of their scalar charges.} and our main goal is to
do so, which, first of all requires a generalization of the definition of
scalar charge of Refs.~\cite{Pacilio:2018gom,Ballesteros:2023iqb}: the
theories considered in Refs.~\cite{Pacilio:2018gom,Ballesteros:2023iqb} have
global symmetries and the scalar charges are related to them. As we are going
to see, the covariant definition can be extended to the theories that we are
going to consider, which have no global symmetries and it can be used to
determine which values are allowed in the presence of a bifurcate black-hole
horizon. As a byproduct we are going to see that the potentials that allow
for asymptotically-flat black-hole solutions with well-defined scalar charges
must satisfy a quite restrictive set of conditions previously found in
Ref.~\cite{Cadoni:2015gfa}.

This paper is organized as follows: in Section~\ref{sec-thetheory} we are
going to describe the kind of theories that we are going to study. In
Section~\ref{sec-scalarcharge} we are going to give a covariant definition of
scalar charge for static solutions of these theories which satisfies a Gauss
law and we are going to see which scalar potentials allow for well-defined
scalar charges in the presence of a bifurcate horizon. In
Section~\ref{sec-smarr} we are going to derive a general Smarr formula for the
black-hole solutions of these theories and in Section~\ref{sec-firstlaw} we
will derive the first law. in Section~\ref{sec-anabalalon-olivaBH} we are
going to test the general results obtained in the previous sections using the
asymptotically-flat Anabal\'on-Oliva black hole \cite{Anabalon:2012ih}. 
The results obtained are discussed in Section~\ref{sec-discussion}, which also
contains pointers to further research.

\section{The theory}
\label{sec-thetheory}

The theory we are going to work with consists of gravity, described by the
Vielbein $e^{a}=e^{a}{}_{\mu}dx^{\mu}$, coupled minimally to a real scalar
field $\phi$ which couples to itself via a scalar potential $V(\phi)$. The
action, in differential-form language, is simply given by\footnote{Our
  conventions are those of Ref.~\cite{Ortin:2015hya,Ballesteros:2023iqb}.}

\begin{equation}
  \label{eq:action}
  \begin{aligned}
    S[e,\phi]
    & =
    \frac{1}{16\pi G_{N}^{(4)}} \int \left\{
      -\star (e^{a}\wedge e^{b})\wedge R_{ab}
      +\tfrac{1}{2}d\phi\wedge \star d\phi +\star V(\phi)
    \right\}
    \\
    & \\
    & \equiv
    \int \mathbf{L}\,,
  \end{aligned}
\end{equation}

\noindent
where $\mathbf{L}$ is the Lagrangian 4-form.

Under a general variation of the fields, the action transforms as

\begin{equation}
  \label{eq:deltaS}
  \delta S
  =
  \int\left\{
    \mathbf{E}_{a}\wedge \delta e^{a} + \mathbf{E}\delta\phi
    +d\mathbf{\Theta}(e,\phi,\delta e,\delta\phi)
    \right\}\,,
\end{equation}

\noindent
where, ignoring the normalization factor $(16\pi G_{N}^{(4)})^{-1}$ for the
time being (we will recover it when necessary), the Einstein equations
$\mathbf{E}_{a}$ and the scalar equation $\mathbf{E}$ are given by

\begin{subequations}
  \begin{align}
    \label{eq:Ea}
    \mathbf{E}_{a}
    & =
      \imath_{a}\star(e^{b}\wedge e^{c})\wedge R_{bc}
      +\tfrac{1}{2}\left(\imath_{a}d\phi\star d\phi
      +d\phi\wedge \imath_{a}\star d\phi\right)
      -\imath_{a}\star V\,,
    \\
    & \nonumber \\
    \label{eq:E}
    \mathbf{E}
    & =
      -d\star d\phi+\star V'\,,
  \end{align}
\end{subequations}

\noindent
where $V'\equiv dV/d\phi$, and where 

\begin{equation}
    \mathbf{\Theta}(e,\phi,\delta e,\delta\phi)
 =
  -\star (e^{a}\wedge e^{b})\wedge \delta \omega_{ab}
  +\star d\phi\delta\phi\,.  
\end{equation}

Several no-hair theorems for this system have been proven in the literature
\cite{Bekenstein:1995un,kn:HeuslerStraumann1992,Heusler:1996ft,Heusler:1992ss,Sudarsky:1995zg}
with the positivity of the scalar potential as one of the main
assumptions. Several asymptotically-flat black-hole solutions with regular
horizon and scalar hair have been found in systems that violate this
particular assumption
\cite{Bechmann:1995sa,Dennhardt:1996cz,Nucamendi:1995ex,Anabalon:2012ih,Cadoni:2015gfa}\footnote{See
  Ref.~\cite{Herdeiro:2015waa} for a review on this topic with many
  references.}. We will only study in detail the asymptotically-flat
Anabal\'on-Oliva black hole Ref.~\cite{Anabalon:2012ih}.

Our goal in this paper is to study how the concept of scalar
charge can be used to study these solutions and what allows them to exist at
all. Thus, we first study the definition of scalar charge in this system.

\section{Scalar charge}
\label{sec-scalarcharge}

Following Refs.~\cite{Pacilio:2018gom,Ballesteros:2023iqb} we take the inner
product of the Killing vector $k$ with the scalar equation of motion, getting

\begin{equation}
  \begin{aligned}
    \imath_{k}\mathbf{E}
    & =
    d\left[\imath_{k}\star d\phi+\mathcal{W}_{k}\right]\,,
  \end{aligned}
\end{equation}

\noindent
where we have defined

\begin{equation}
  \label{eq:Wkdefinition}
  d\mathcal{W}_{k}
  =
  \imath_{k}\star V'\,.
\end{equation}

The existence of the 2-form $\mathcal{W}_{k}$ is (locally) guaranteed by the
assumptions concerning the symmetry of the system: if the diffeomorphism
generated by $k$ leaves invariant all the fields of the configuration,

\begin{equation}
\pounds_{k}\star V' = d\imath_{k}\star V' =0\,.   
\end{equation}

\noindent
Then, we define the scalar charge 2-form associated to the Killing vector $k$,
$\mathbf{Q}_{\phi}[k]$, by\footnote{The sign and normalization have been
  chosen so as to reproduce the conventional value of the scalar charge in
  absence of a potential. This is defined by the asymptotic expansion
  Eq.~(\ref{eq:asymptoticdefinition}).}

\begin{equation}
  \mathbf{Q}_{\phi}[k]
  \equiv
  -\frac{1}{4\pi G_{N}^{(4)}}
  \left[\imath_{k}\star d\phi+\mathcal{W}_{k}\right]\,.  
\end{equation}

\noindent
We have just shown that $ \mathbf{Q}_{\phi}[k]$ is closed on-shell or, in
other words, that it satisfies a Gauss law on-shell.

We are going to consider stationary black-hole spacetimes and $k$ will be the
timelike Killing vector that generates their Killing horizon. As we will show
later, for spherically-symmetric, static, asymptotically-flat black holes,
this choice gives the scalar charge $\Sigma$ defined in
Eq.~(\ref{eq:asymptoticdefinition}) as the integral over any closed
2-dimensional surface $\Sigma^{2}$ enclosing the black hole horizon

\begin{equation}
  \label{eq:Sigmadef}
  \Sigma
  \equiv
  \int_{\Sigma^{2}}\mathbf{Q}_{\phi}[k]\,.
\end{equation}

The on-shell closedness of $\mathbf{Q}_{\phi}[k]$ ensures that this definition
does not depend on the integration surface chosen as long as they are
homologically equivalent.

$\mathcal{W}_{k}$ is defined up to closed forms. We can use that
freedom to make it vanish at spatial infinity:

\begin{equation}
  \label{eq:bc}
  \mathcal{W}_{k}(\infty)
  =
  0\,.
\end{equation}

\noindent
Then, if we integrate $\mathbf{Q}_{\phi}[k]$ over the 2-sphere at spatial
infinity, $S^{2}_{\infty}$, we find that

\begin{equation}
  \label{eq:Sigmaversusphi}
  \Sigma
  =
  -\frac{1}{4\pi G_{N}^{(4)}} \int_{S^{2}_{\infty}}\imath_{k}\star d\phi\,,
\end{equation}

\noindent
which recovers the conventional definition of scalar charge, as we are going
to see.

If the black hole has a bifurcate horizon and we choose to integrate
$\mathbf{Q}_{\phi}[k]$ over the bifurcation surface $\mathcal{BH}$ in which
$k=0$, we get

\begin{equation}
  \Sigma
  =
  -\frac{1}{4\pi G_{N}^{(4)}}\int_{\mathcal{BH}}\mathcal{W}_{k}\,,
\end{equation}

\noindent
which provides an interesting relation between the scalar potential and the
scalar charge of a black hole with bifurcate horizon. If we use the boundary
condition Eq.~(\ref{eq:bc}) and the definition of $\mathcal{W}_{k}$
Eq.~(\ref{eq:Wkdefinition}), applying Stokes' theorem we can rewrite the above
formula in the form

\begin{equation}
  \label{eq:SigmaversusV'}
  \Sigma
  =
  -\frac{1}{4\pi G_{N}^{(4)}}\int_{\Sigma^{3}}\imath_{k}\star V'\,,
\end{equation}

\noindent
where $\Sigma^{3}$ is a hypersurface with boundaries at the bifurcation
surface and spatial infinity.

Black holes with regular bifurcate horizons and scalar hair corresponding to
the scalar charge $\Sigma$ will only exist if the integral in the right-hand
side is finite, which imposes strong conditions on the scalar potentials that
allow for hairy black hole solutions. In order to find these conditions, we
are going to focus on static, asymptotically-flat, spherically-symmetric black
holes with metrics of the form

\begin{equation}
  \label{eq:sphericallysymmetricmetric}
  ds^{2}
  =
  \lambda (\rho)dt^{2}-\lambda^{-1}(\rho)dr^{2}-R^{2}(\rho)d\Omega^{2}_{(2)}\,.
\end{equation}

Since the integral of $\mathbf{Q}_{\phi}[k]$ over any 2-sphere of constant
radius $\rho$ should give the same result, $\Sigma$, 

\begin{equation}
  \label{eq:QkSigma}
  \mathbf{Q}_{\phi}[k]
  =
  \frac{\Sigma}{16\pi}  \omega_{(2)}\,,
\end{equation}

\noindent
where $\omega_{(2)}$ is the volume 2-form of the unit sphere.

On the other hand, 

\begin{equation}
  \label{eq:firsttermQphi}
  \begin{aligned}
    -\imath_{k}\star d\phi
    & =
    -\lambda R^{2}\partial_{\rho}\phi \omega_{(2)}\,, 
  \end{aligned}
\end{equation}

\noindent
and, if $\phi$ behaves at spatial infinity as\footnote{This is the
  conventional definition of the scalar charge $\Sigma$.}

\begin{equation}
  \label{eq:asymptoticdefinition}
\phi \sim \phi_{\infty} +\frac{G_{N}^{(4)}\Sigma}{\rho}+\mathcal{O}(\rho^{-2})\,,  
\end{equation}

\noindent
where $\phi_{\infty}$ is the constant value of the scalar at spatial infinity,
we find that, in that limit,

\begin{equation}
  \label{eq:istardphi}
  -\frac{1}{4\pi G_{N}^{(4)}}  \imath_{k}\star d\phi
  \sim
  \frac{\Sigma}{4 \pi}  \omega_{(2)} +\mathcal{O}(\rho^{-1})\,.
\end{equation}

\noindent
which implies that, in the same limit, 

\begin{equation}
  \mathcal{W}_{k}
  \sim \mathcal{O}(\rho^{-1})\,,  
\end{equation}

\noindent
which is consistent with the boundary condition we had chosen for
$\mathcal{W}_{k}$, Eq.~(\ref{eq:bc}), and, in its turn, implies that

\begin{equation}
-\frac{1}{4\pi G_{N}^{(4)}}
\mathcal{W}_{k}(\mathcal{BH})
=
\frac{\Sigma}{4\pi}  \omega_{(2)}\,.  
\end{equation}

Observe that, since $\lambda$ must vanish on the horizon,
$\imath_{k}\star d\phi$ vanishes everywhere on the horizon and not just on the
bifurcation surface.

Let us find an explicit expression for $\mathcal{W}_{k}$. For the metrics we
are dealing with,

\begin{equation}
  \label{eq:Wk}
  \begin{aligned}
    d\mathcal{W}_{k}
    & =
d\left[\mathcal{W}_{k}(\rho)\omega_{(2)}\right]\,,
  \end{aligned}
\end{equation}

\noindent
where the function $\mathcal{W}_{k}(\rho)$ is defined by

\begin{equation}
  \label{eq:Wkr}
  d\mathcal{W}_{k}(\rho)
  =
  -R^{2}(\rho) V'[\phi(\rho)] d\rho\,.
\end{equation}

\noindent
From Eq.~(\ref{eq:QkSigma}) and  the definition of $\mathbf{Q}_{\phi}[k]$

\begin{equation}
-\frac{1}{4\pi G_{N}^{(4)}}
\left[\imath_{k}\star d\phi+\mathcal{W}_{k}\right]
=
\frac{\Sigma}{4\pi}  \omega_{(2)}\,,
\end{equation}

\noindent
and, using Eqs.~(\ref{eq:firsttermQphi}) and (\ref{eq:Wk}) we have

\begin{equation}
  \label{eq:esa}
\mathcal{W}_{k}(\rho)
=
-\Sigma G_{N}^{(4)}-R^{2}\partial_{\rho}\phi
=
\int_{\rho}^{\infty} R^{2}(\rho) V'[\phi(\rho)]\,.
\end{equation}

If we expand asymptotically the right-hand side of the definition of
$\mathcal{W}_{k}(\rho)$, Eq.~(\ref{eq:Wkr}), assuming

\begin{equation}
  \phi \sim \phi_{\infty} +\frac{G_{N}^{(4)}\Sigma}{r}+ \frac{\Delta}{r^{2}}
  +\mathcal{O}(r^{-3})\,,
\end{equation}

\noindent
we get

\begin{equation}
  \label{eq:Wkrexpansion}
  \begin{aligned}
    d\mathcal{W}_{k}(r)
    & =
    \left(1-\frac{2(G_{N}^{(4)}M)^{2}}{r^{2}}
      +\mathcal{O}(r^{-3})\right) \times
    \\
    & \\
    & \hspace{.5cm} \times\left\{ V'[\phi_{\infty}] r^{2}
      +V''[\phi_{\infty}]\Sigma G_{N}^{(4)} r
      +\tfrac{1}{2}\left[V'''[\phi_{\infty}]\left(\Sigma
          G_{N}^{(4)}\right)^{2}
        +2V''[\phi_{\infty}]\Delta\right]+\mathcal{O}(r^{-3}) \right\} dr
    \\
    & \\
    & =
    \left(1-\frac{2(G_{N}^{(4)}M)^{2}}{r^{2}} +\mathcal{O}(r^{-3})\right)
    \times
    \\
    & \\
    & \hspace{.5cm}
    \times \left\{ V'[\phi_{\infty}] r^{2}
      +V''[\phi_{\infty}]\Sigma G_{N}^{(4)} r
      +\tfrac{1}{2}V'''[\phi_{\infty}]\left(\Sigma G_{N}^{(4)}\right)^{2}
      +V''[\phi_{\infty}]\Delta \right.
    \\
    & \\
    & \hspace{.5cm} \left.  -2(G_{N}^{(4)}M)^{2}V'[\phi_{\infty}]
      +\mathcal{O}(r^{-1}) \right\}dr\,,
  \end{aligned}
\end{equation}

\noindent
and, comparing with an asymptotic expansion of $\mathcal{W}_{k}(r)$ that takes
into account that $\mathcal{W}_{k}(\infty)=0$

\begin{equation}
  \mathcal{W}_{k}(r)
  =
  \mathcal{O}(r^{-1})\,,
  \,\,\,\,\,
  \Rightarrow
  \,\,\,\,\,
  d\mathcal{W}_{k}(r)
  =
  \mathcal{O}(r^{-2})dr\,,
\end{equation}

\noindent
we find that the potential and its first four derivatives must vanish at the
asymptotic value of the scalar:

\begin{equation}
  \label{eq:conditions}
  V(\phi_{\infty})
  =
  V'[\phi_{\infty}]
  =
  V''[\phi_{\infty}]
  =
  V'''[\phi_{\infty}]
  =
  V''''[\phi_{\infty}]
  =
  0\,,
\end{equation}

\noindent
where we have taken into account that we are considering asymptotically-flat
black holes only and where we have assumed that $\Sigma\neq 0$.\footnote{These
  conditions have been previously derived in a different but equivalent way in
  Ref.~\cite{Cadoni:2015gfa} in which the requirement of a well-defined scalar
  charge has been implicitly used.}

These conditions are satisfied by the potential of the theory of
Ref.~\cite{Anabalon:2012ih} for the asymptotic value of the
asymptotically-flat (Anabal\'on-Oliva) black hole
$\phi_{\infty}=0$.\footnote{They are also satisfied by the first scalar
  potential of Ref.~\cite{Dennhardt:1996cz}. The scalar potential of
  Ref.~\cite{Bechmann:1995sa} was not given in full. On the other hand, the
  scalar potential in Ref.~\cite{Nucamendi:1995ex} manifestly violates those
  conditions, but the asymptotic behaviour of the scalar field is exponential,
  $\phi \sim e^{-\alpha \rho}/\rho$ so that $\Sigma=0$.}  They are not
satisfied for a massive scalar, though, because
$ V''[\phi_{\infty}]=m^{2}\neq 0$. Therefore, the result that we have
obtained, based on a definition of scalar charge that satisfies a Gauss law is
equivalent to Bekenstein's no-hair theorem of Ref.~\cite{Bekenstein:1972ny}
and also discards many other scalar potentials.

\section{Smarr formula}
\label{sec-smarr}

Our next step in the study of these theories is the derivation of a Smarr
formula using the techniques developed in
Refs.~\cite{Kastor:2008xb,Kastor:2010gq,Liberati:2015xcp,Hajian:2021hje,Ortin:2021ade,Mitsios:2021zrn,Meessen:2022hcg}.

The Smarr formula follows from the integration of the generalized Komar charge
\cite{Komar:1958wp} on the hypersurface $\Sigma^{3}$ that interpolates
between the bifurcation sphere and the sphere at spatial infinity (its two
boundaries). The generalized Komar charge is given by

\begin{equation}
  \mathbf{K}[k]
\equiv
-\left(\mathbf{Q}[k]-\omega_{k}\right)\,,
\end{equation}

\noindent
where $\mathbf{Q}[k]$ is the Noether-Wald charge associated to the Killing
vector $k$ and $\omega_{k}$ is the 2-form implicitly defined by\footnote{We
  indicate relations which only hold on-shell with $\doteq$.}

\begin{equation}
  \label{eq:omegak}
\imath_{k}\mathbf{L}
\doteq
d\omega_{k}\,.
\end{equation}

The (local) existence of $\omega_{k}$, as that of $\mathcal{W}_{k}$, is
guaranteed by the assumption that $k$ generates a symmetry of all the field of
the solution:

\begin{equation}
  \pounds_{k}\mathbf{L}
  =
  d\imath_{k}\mathbf{L}
  =
  0\,.
\end{equation}

As explained in Ref.~\cite{Elgood:2020svt}, in order to compute the
Noether-Wald charge one must properly take into account the gauge freedoms of
the fields of the theory. In this case, the only gauge symmetry of the theory
(apart from diffeomorphisms), is the local Lorentz symmetry acting on the
Vielbein and the right way to deal with it is to replace the Lie derivative of
the Vielbein by the Lorentz-covariant (or Lie-Lorentz) derivative
\cite{Ortin:2002qb} (see also Ref.~\cite{Jacobson:2015uqa}), so that

\begin{equation}
  \label{eq:deltaxitransformations}
  \delta_{\xi}e^{a}
  =
  -\left(\mathcal{D}\xi^{a}+P_{\xi}{}^{a}{}_{b}e^{b}\right)\,,
  \hspace{1cm}
  \delta_{\xi}\phi
  =
  -\imath_{\xi}d\phi\,,
\end{equation}

\noindent
where $P_{\xi}{}^{ab}$, the \textit{Lorentz momentum map}, is defined by the
equation

\begin{equation}
  \mathcal{D}P_{k}{}^{ab}+\imath_{k}R^{ab}
  =
  0\,,
\end{equation}

\noindent
for Killing vectors $k$. This equation is satisfied by the \textit{Killing
  bivector}

\begin{equation}
  P_{k}{}^{ab}
  =
  \nabla^{a}k^{b}\,.  
\end{equation}

Substituting the transformations Eqs.~(\ref{eq:deltaxitransformations}) into
Eq.~(\ref{eq:deltaS}) and using the Noether identity associated to the
invariance under local Lorentz transformations of the action

\begin{equation}
  \begin{aligned}
    P_{\xi}{}^{a}{}_{b}\mathbf{E}_{a}\wedge e^{b}
    & =
      0\,,
  \end{aligned}
\end{equation}

\noindent
and the Noether identity associated to the invariance under diffeomorphisms of
the action 





\begin{equation}
  \begin{aligned}
   \mathcal{D}\mathbf{E}_{a}
      +\mathbf{E}\imath_{a}d\phi
      & =
  0\,,
  \end{aligned}
\end{equation}

\noindent
we find that the variation of the action under the transformations
Eqs.~(\ref{eq:deltaxitransformations}) 

\begin{equation}
  \begin{aligned}
    \delta_{\xi} S
    & =
    -\int\left\{ \left(\mathcal{D}\mathbf{E}_{a}
       +\mathbf{E}\imath_{a}d\phi\right)\xi^{a}
      +P_{\xi}{}^{a}{}_{b}\mathbf{E}_{a}\wedge e^{b}
    \right.
    \\
    & \\
    & \hspace{.5cm}
    \left.
      +d\left[-\mathbf{E}_{a}\xi^{a}
        -\star (e^{a}\wedge e^{b})\wedge
        \left(\imath_{\xi} R_{ab} +\mathcal{D}P_{\xi\, ab}\right)
  +\star d\phi \imath_{\xi}d\phi\right]
    \right\}\,,
  \end{aligned}
\end{equation}

\noindent
is just a total derivative.

Massaging this total derivative a bit, we arrive to 

\begin{equation}
  \begin{aligned}
    \delta_{\xi} S
    & =
    -\int d\left\{\imath_{\xi}\mathbf{L}
        -d\left[\star (e^{a}\wedge e^{b})P_{\xi\, ab}\right]
       \right\}\,.
  \end{aligned}
\end{equation}

Since the action is only invariant under a total derivative under these
transformations, we arrive to the Noether-Wald charge of pure Einstein gravity
\cite{Wald:1993nt}, which is nothing but the Komar charge of pure Einstein
gravity

\begin{equation}
  \label{eq:NoetherWald}
  \mathbf{Q}[\xi]
  =
  \frac{1}{16\pi G_{N}^{(4)}}\star (e^{a}\wedge e^{b})P_{\xi\, ab}\,.  
\end{equation}

Now, in order to find $\omega_{k}$ we need to evaluate the on-shell
Lagrangian. We first take the trace of the Einstein equation

\begin{equation}
  \begin{aligned}
    e^{a}\wedge \mathbf{E}_{a}
    & =
    -2\left[\mathbf{L} +\star V\right]\,.
  \end{aligned}
\end{equation}

\noindent
Then,

\begin{equation}
  \mathbf{L}
  =
  -\tfrac{1}{2} e^{a}\wedge \mathbf{E}_{a} -\star V
  \doteq
  -\star V\,,
\end{equation}

\noindent
and, for a Killing vector $k$ that leaves all the fields invariant 

\begin{equation}
  \label{eq:Vkdef}
  \imath_{k}\mathbf{L}
  \doteq
  -\imath_{k}\star V
  \equiv
  d\mathcal{V}_{k}\,.
\end{equation}

Again, the (local) existence of the 2-form $\mathcal{V}_{k}$ is guaranteed by
the assumptions on the symmetry of the configurations.

The Komar charge of this theory is finally given by

\begin{equation}
  \label{eq:Komarcharge}
  \mathbf{K}[k]
  =
  -\frac{1}{16\pi G_{N}^{(4)}}\left[\star (e^{a}\wedge e^{b})P_{k\, ab}
    -\mathcal{V}_{k} \right]\,,
\end{equation}

\noindent
and it is not difficult to check that it is closed on-shell:


\begin{equation}
  d\mathbf{K}[k]
  =
    \tfrac{1}{2} e^{a}\wedge \mathbf{E}_{a}
  -k^{a}\mathbf{E}_{a}
  +\frac{1}{16\pi G_{N}^{(4)}}\imath_{k}d\phi \star d\phi
  \doteq
  0\,,
\end{equation}

\noindent
recalling that, by assumption, $\imath_{k}d\phi=0$.

Let us consider asymptotically-flat ($V=0$ at spatial infinity), static black
holes with bifurcate Killing horizons $\mathcal{H}$ associated to $k$ and let
us integrate $d\mathbf{K}[k]$ over a hypersurface $\Sigma^{3}$ whose
boundaries are the bifurcation sphere $\mathcal{BH}$ where $k=0$ and the
2-sphere at spatial infinity S$^{2}_{\infty}$. Applying Stokes theorem

\begin{equation}
  0
  \doteq
\int_{\mathrm{S}^{2}_{\infty}}\mathbf{K}[k]
  -\int_{\mathcal{BH}}\mathbf{K}[k]\,.
\end{equation}

\noindent
At infinity, by assumption

\begin{equation}
  \int_{\mathrm{S}^{2}_{\infty}}\mathbf{K}[k]
  =
  \tfrac{1}{2}M
  +\frac{1}{16\pi G_{N}^{(4)}} \int_{\mathrm{S}^{2}_{\infty}}\mathcal{V}_{k}\,.
\end{equation}

\noindent
Over the bifurcation sphere

\begin{equation}
  \int_{\mathcal{BH}}\mathbf{K}[k]
  =
  \frac{\kappa A}{4 G_{N}^{(4)}}
  +\frac{1}{16\pi G_{N}^{(4)}} \int_{\mathcal{BH}}\mathcal{V}_{k}\,.
\end{equation}

\noindent
Thus, using again Stokes' theorem and the definition of $\mathcal{V}_{k}$
Eq.~(\ref{eq:Vkdef}), we arrive at the Smarr formula

\begin{equation}
  \label{eq:Smarrformula1}
  \begin{aligned}
    M
    & =
    2ST -\frac{1}{8\pi G_{N}^{(4)}}
    \left[
    \int_{\mathrm{S}^{2}_{\infty}}\mathcal{V}_{k}
-\int_{\mathcal{BH}}\mathcal{V}_{k}
     \right]
    \\
    & \\
    & =
    2ST -\frac{1}{8\pi G_{N}^{(4)}}
  \int_{\Sigma^{3}}\imath_{k}\star V\,.
  \end{aligned}
\end{equation}

This is not the form in which the Smarr formula is usually presented. The
scalar potential must be proportional to one or several dimensionful coupling
constants. Let $\alpha$ be that constant and let it have dimensions of inverse
length squared so that

\begin{equation}
V = \alpha \frac{\partial V}{\partial\alpha}\,.
\end{equation}

\noindent
Then, the Smarr formula can be written in the more standard form 

\begin{equation}
  \label{eq:Smarrformula2}
    M
     =
     2ST +2\alpha \Phi_{\alpha}\,,
     \,\,\,\,\,
     \text{where}
     \,\,\,\,\,
     \Phi_{\alpha}
     \equiv
     -\frac{1}{16\pi G_{N}^{(4)}}
  \int_{\Sigma^{3}}\imath_{k}\star \frac{\partial V}{\partial \alpha}\,,
\end{equation}

\noindent
which, as proposed in
Refs.~\cite{Hajian:2021hje,Ortin:2021ade,Meessen:2022hcg}, must be interpreted
in terms of extended thermodynamics: $\alpha$ plays the role of a new
thermodynamic variable and $\Phi_{\alpha}$ plays the role of its conjugate
potential. The validity of this Smarr formula can be tested directly in the
existing hairy solutions and we will do so for the Anabal\'on-Oliva black hole.

For the static, spherically symmetric metrics we are considering

\begin{equation}
  \begin{aligned}
    \imath_{k}\star V
    & =
    -R^{2}(\rho)V(\phi(\rho))d\rho \wedge \omega_{(2)}\,,
  \end{aligned}
\end{equation}

\noindent
and

\begin{equation}
  \label{eq:Phialpha}
  \Phi_{\alpha}
  =
      \frac{1}{4G_{N}^{(4)}}
  \int^{\infty}_{\rho_{h}} R^{2}\frac{\partial V}{\partial \alpha}d\rho\,.
\end{equation}

\noindent
For large values of $\rho$

\begin{equation}
  \begin{aligned}
    \imath_{k}\star V
    & =
    -\left(1-\frac{2(G_{N}^{(4)}M)^{2}}{\rho^{2}}
      +\mathcal{O}(\rho^{-3})\right)
    \times
    \\
    & \\
    & \hspace{.5cm}
    \times \rho^{2}\left\{
      V(\phi_{\infty})
    +V'(\phi_{\infty})\frac{\Sigma G_{N}^{(4)}}{\rho}+\cdots
\right\}d\rho \wedge \omega_{(2)}\,.  
  \end{aligned}
\end{equation}

Asymptotic flatness implies $V(\phi_{\infty})=0$ and the convergence of the
integral demands, again, Eqs.~(\ref{eq:conditions}) to hold.

\section{The first  law of black hole mechanics}
\label{sec-firstlaw}

As shown in Ref.~\cite{Meessen:2022hcg}, in order to derive the first law of
black-hole mechanics using Wald's formalism in theories with dimensionful
parameters such as those that must necessarily occur in the scalar potential
($\alpha$ in the case discussed in Section~\ref{sec-smarr}), it is necessary
to dualize those parameters into $(d-1)$-form potentials which have a gauge
symmetry generated by $(d-2)$-form parameters and work with the dual
formulation of the theory. In particular, we have to rederive the Noether-Wald
charge, which will have an additional term associated to the new gauge
symmetry.

\subsection{The dual theory}

We can directly draw from the results of Ref.~\cite{Meessen:2022hcg} and write
an action which contains two additional dynamical fields: the scalar
$\vartheta$ and the 3-form $C$. $\vartheta$ is the square root of the
dimensionful constant $\alpha$ discussed in Section~\ref{sec-smarr}, which in
this setting is promoted to a scalar field,

\begin{equation}
  \alpha
  =
  \vartheta^{2}\,,  
\end{equation}

\noindent
so that

\begin{equation}
  \vartheta \frac{\partial V}{\partial\vartheta}
  =
  2V\,.
\end{equation}

\noindent
The 3-form $C$ is the dual of $\vartheta$ and it is introduced in the action
as a Lagrange multiplier enforcing the constraint $d\vartheta=0$. The action
is that in Eq.~(\ref{eq:action}) supplemented by the Lagrange multiplier term,
which is topological and does not modify the Einstein equations:

\begin{equation}
  \label{eq:dualaction}
  \begin{aligned}
    S[e,\phi,\vartheta,C]
    & =
    \frac{1}{16\pi G_{N}^{(4)}} \int \left\{
      -\star (e^{a}\wedge e^{b})\wedge R_{ab}
      +\tfrac{1}{2}d\phi\wedge \star d\phi
      -C\wedge d\vartheta +\star V(\phi)
    \right\}
    \\
    & \\
    & \equiv
    \int \mathbf{L}\,.
  \end{aligned}
\end{equation}

Under a general variation of the fields, 

\begin{equation}
  \label{eq:generalvariationdualaction}
  \delta S
 =
  \int \left\{
    \mathbf{E}_{a}\wedge \delta e^{a}
    +\mathbf{E}\delta\phi
    +\mathbf{E}_{C}\wedge \delta C
    +\mathbf{E}_{\vartheta}\wedge \delta \vartheta
     +d\mathbf{\Theta}(\varphi,\delta\varphi)
  \right\}\,,
\end{equation}

\noindent
where the Einstein and scalar equations $\mathbf{E}_{a},\mathbf{E}$ are
identical to those of the original action Eqs.~(\ref{eq:Ea}) and (\ref{eq:E}),
and the equations of the 3-form $C$ and of the dimensionful ``constant''
$\vartheta$ are given by

\begin{subequations}
  \begin{align}
        \mathbf{E}_{\vartheta}
    & =
     -G +\star \frac{\partial V}{\partial \vartheta}\,,
    \\
    & \nonumber \\
        \mathbf{E}_{C}
    & =
      d\vartheta\,,
  \end{align}
\end{subequations}

\noindent
where

\begin{equation}
  G
  \equiv
  dC\,,
\end{equation}

\noindent
is the field strength of the 3-form $C$, invariant under the gauge
transformations

\begin{equation}
  \delta_{\chi}C
  =
  d\chi\,,
\end{equation}

\noindent
where $\chi$ is an arbitrary 2-form.

On-shell, the equation of motion of $\vartheta$ is a duality relation between
the 3-form and $\vartheta$ \cite{Bergshoeff:2009ph}

\begin{equation}
  G
  =
  \star \frac{\partial V}{\partial \vartheta}\,,  
\end{equation}

\noindent
and, as expected, the equation of motion of $C$ just says that $\vartheta$ is
constant.\footnote{In presence of sources of 3-form $C$, 2-branes (domain
  walls) charged with respect to $C$, it can be piecewise constant, see for
  instance, \cite{Bergshoeff:2001pv}. This action, thus, is slightly more
  general than the original one.}

Finally, $\mathbf{\Theta}$ contains a new term

\begin{equation}
  \label{eq:Thetadual}
    \mathbf{\Theta}(\varphi,\delta\varphi)
    =
    -\star (e^{a}\wedge e^{b})\wedge \delta \omega_{ab}
    +\star d\phi\delta\phi
    +C\delta\vartheta\,.
\end{equation}

Observe that the action Eq.~(\ref{eq:dualaction}) is only invariant under the
gauge transformations of $C$ up to a total derivative (defined itself up to
total derivatives), which we will have to take into account:

\begin{equation}
  \label{eq:totalderivative}
  \delta_{\chi}S
  =
 \int d\left(\vartheta d\chi \right)\,.
\end{equation}

\subsection{The Noether-Wald charge of the dual theory}

We just have to use the general form of the variation of the dual action
 Eq.~(\ref{eq:generalvariationdualaction}) in the particular case of the
 variations Eqs.~(\ref{eq:deltaxitransformations}) and

\begin{equation}
  \label{eq:deltaxitransformationsdual}
  \delta_{\xi}\vartheta
  =
  -\imath_{\xi}\vartheta\,,
  \hspace{1cm}
  \delta_{\xi}C
  =
  -\left(\imath_{\xi}G +dP_{\xi}\right)\,,
\end{equation}

\noindent
where, following the general rules, we have defined the momentum map 2-form 
$P_{\xi}$ to satisfy the equation

\begin{equation}
  \label{eq:momentummapeqG}
  \imath_{k}G +dP_{k}
  =
  0\,,
\end{equation}

\noindent
for a Killing vector $k$ that leaves invariant all the fields of the
theory. This assumption guarantees the existence of $P_{k}$.

The Noether-Wald and Komar charges can be simply read from
Ref.~\cite{Meessen:2022hcg} 

\begin{subequations}
  \begin{align}
  \label{eq:NoetherWalddual}
  \mathbf{Q}[\xi]
  & =
    \frac{1}{16\pi G_{N}^{(4)}}\left[\star (e^{a}\wedge e^{b})P_{\xi\, ab}
    +\vartheta P_{\xi} \right]\,,
    \\
    & \nonumber \\
    \label{eq:Komarchargedual}
  \mathbf{K}[\xi]
  & =
  -\frac{1}{16\pi G_{N}^{(4)}}\left[\star (e^{a}\wedge e^{b})P_{\xi\, ab}
    -\tfrac{1}{2}\vartheta P_{\xi}\right]\,,
  \end{align}
\end{subequations}

\noindent
but we have to take into account that they have been computed using the
particular choice of total derivative Eq.~(\ref{eq:totalderivative})

\begin{equation}
  \label{eq:choiceoftotalderivative}
  \delta_{\xi}S
  =
  \int \left\{-d\vartheta\wedge \imath_{\xi}C - \vartheta dP_{\xi}\right\}\,.
\end{equation}

Before deriving the Smarr formula and the first law for the dual theory, it is
necessary to consider the generalized, restricted, zeroth law
\cite{Elgood:2020svt}. If $k$ is the Killing vector that generates the
horizon, in the bifurcation surface where $k=0$ Eq.~(\ref{eq:momentummapeqG})
implies that

\begin{equation}
  dP_{k}
  \stackrel{\mathcal{BH}}{=}
  0\,.
\end{equation}

In the static case in which we are interested here, this just means that

\begin{equation}
  P_{k}
  \stackrel{\mathcal{BH}}{=}
  16\pi G_{N}^{(4)}\Phi_{G} \omega_{(2)}\,,
\end{equation}

\noindent
where $\Phi_{G}$ is a constant and $\omega_{(2)}$ is the volume form of the
unit 2-sphere.

Observe that, on-shell and using the homogeneity of $V$ in $\vartheta$, the
equation of motion of $C$, the momentum map equation (\ref{eq:momentummapeqG})
and the equation of motion of $\vartheta$ we have the following relation:

\begin{equation}
  \imath_{k}\star V
  =
  \frac{1}{2}\vartheta
  \imath_{k}\star \frac{\partial V}{\partial \vartheta}
  \doteq
  \frac{1}{2}\vartheta  \imath_{k}G
  =
  -\frac{1}{2}\vartheta d P_{k}
  \doteq
  d\left(-\frac{1}{2}\vartheta P_{k}\right)\,,
\end{equation}

\noindent
which means that $\mathcal{V}_{k}$ defined in Eq.~(\ref{eq:Vkdef}) is given by 

\begin{equation}
  \label{eq:VkvarthetaP}
  \tfrac{1}{2}\vartheta P_{k}
  \doteq
  \mathcal{V}_{k}\,.
\end{equation}

Replacing this result in the Komar charge Eq.~(\ref{eq:Komarchargedual}) we
recover that of the original theory Eq.~(\ref{eq:Komarcharge}), which means
that we get the same Smarr formula Eq.~(\ref{eq:Smarrformula2}).

Let us now consider the derivation of the first law in full detail, improving
the derivations made in
Refs.~\cite{Elgood:2020svt,Elgood:2020mdx,Elgood:2020nls,Mitsios:2021zrn,Meessen:2022hcg,Ortin:2022uxa,Ballesteros:2023iqb}
in which either scalar charges had not been considered or the action was
exactly invariant under gauge transformations (which is not the case here).

We start by defining the symplectic 3-form
\cite{Lee:1990nz,Wald:1993nt,Iyer:1994ys}

\begin{equation}
  \label{eq:symplectic3form}
  \omega(\varphi,\delta_{1}\varphi,\delta_{2}\varphi)
  \equiv
  \delta_{1}\mathbf{\Theta}(\varphi,\delta_{2}\varphi)
-\delta_{2}\mathbf{\Theta}(\varphi,\delta_{1}\varphi)\,,
\end{equation}

\noindent
where $\varphi$ denotes collectively all the fields of the theory. In this
case we have to choose in Eq.~(\ref{eq:symplectic3form})
$\delta_{1}\varphi=\delta\varphi$, variations of the fields which satisfy the
linearized equations of motion but which are, otherwise, arbitrary, and
$\delta_{2}\varphi=\delta_{\xi}\varphi$, the transformations under
diffeomorphisms given in Eqs.~(\ref{eq:deltaxitransformations}) and
(\ref{eq:deltaxitransformationsdual}) which, we must recall, include induced
gauge transformations $\delta_{\sigma_{\xi}},\delta_{\chi_{\xi}}$ which we
will denote, collectively, by $\delta_{\Lambda_{\xi}}$, so that

\begin{equation}
\delta_{\xi} = -\pounds_{\xi} +\delta_{\Lambda_{\xi}}\,.
\end{equation}

Under these transformations, using the general expression for the variation of
the action, we get

\begin{equation}
  \label{eq:thetaprime}
  \delta_{\xi}S
  =
  \int d\mathbf{\Theta}'(\varphi,\delta_{\xi}\varphi)\,,
\end{equation}

\noindent
where $\mathbf{\Theta}'$ is a combination of $\mathbf{\Theta}$ and equations
of motion, so that, on-shell, $\mathbf{\Theta}=\mathbf{\Theta}'$.

On the other hand, varying directly the action we find that

\begin{equation}
  \label{eq:theothertotalderivative}
  \delta_{\xi}S
  =
  \int d\left\{-\imath_{\xi}\mathbf{L} +\mathbf{X}(\delta_{\Lambda_{\xi}}\varphi) \right\}\,,
  \,\,\,\,\,
  \text{where}
  \,\,\,\,\,
  d\mathbf{X}(\delta_{\Lambda_{\xi}}\varphi)
  \equiv
  \delta_{\Lambda_{\xi}}\mathbf{L}\,.
\end{equation}

In the theory we are considering we can read $X(\Lambda_{\xi})$ from
Eq.~(\ref{eq:choiceoftotalderivative}):

\begin{equation}
  \label{eq:X}
  \mathbf{X}(\delta_{\Lambda_{\xi}}\varphi)
=
  -d\vartheta\wedge \imath_{\xi}C - \vartheta dP_{\xi}
  \doteq
  d\left(- \vartheta P_{\xi}\right)\,.
\end{equation}

Equating Eqs.~(\ref{eq:thetaprime}) and (\ref{eq:theothertotalderivative}) we
arrive at

\begin{equation}
  \label{eq:Jxidef}
  d\mathbf{J}[\xi]
  =
  0\,,
  \,\,\,\,\,
  \text{where}
  \,\,\,\,\,
  \mathbf{J}[\xi]
  \equiv
  \mathbf{\Theta}'(\varphi,\delta_{\xi}\varphi)
  +\imath_{\xi}\mathbf{L} -\mathbf{X}(\delta_{\Lambda_{\xi}}\varphi)\,.
\end{equation}

Then, 

\begin{equation}
  \begin{aligned}
    \omega(\varphi,\delta\varphi,\delta_{\xi}\varphi)
    & \doteq
    \delta\mathbf{\Theta}'(\varphi,\delta_{\xi}\varphi)
    -\delta_{\xi}\mathbf{\Theta}'(\varphi,\delta\varphi)
    \\
    & \\
    & =
    \delta\left(\mathbf{J}[\xi]-\imath_{\xi}\mathbf{L}+\mathbf{X}(\delta_{\Lambda_{\xi}}\varphi)\right)
    -\left(-\pounds_{\xi}+\delta_{\Lambda_{\xi}}\right)
    \mathbf{\Theta}'(\varphi,\delta\varphi)
    \\
    & \\
    & =
    \delta\mathbf{J}[\xi]
    -\imath_{\xi}\delta\mathbf{L}
    +\delta \mathbf{X}(\delta_{\Lambda_{\xi}}\varphi)
    +\left(\imath_{\xi}d+d\imath_{\xi}-\delta_{\Lambda_{\xi}}\right)
    \mathbf{\Theta}'(\varphi,\delta\varphi)
    \\
    & \\
    & =
    \delta d\mathbf{Q}[\xi]
    -\imath_{\xi}\left(\mathbf{E}_{\varphi}\wedge \delta\varphi
      +d \mathbf{\Theta}(\varphi,\delta\varphi)\right)
    \\
    & \\
    & \hspace{.5cm}
    +\left(\imath_{\xi}d+d\imath_{\xi}-\delta_{\Lambda_{\xi}}\right)
    \mathbf{\Theta}'(\varphi,\delta\varphi)
    +\delta \mathbf{X}(\delta_{\Lambda_{\xi}}\varphi)
    \\
    & \\
    & \doteq
    d\left[\delta \mathbf{Q}[\xi]
    +\imath_{\xi} \mathbf{\Theta}'(\varphi,\delta\varphi)\right]
    -\delta_{\Lambda_{\xi}}\mathbf{\Theta}'(\varphi,\delta\varphi)
    +\delta \mathbf{X}(\delta_{\Lambda_{\xi}}\varphi)\,.
  \end{aligned}
\end{equation}

The last two terms must also be total derivatives:

\begin{subequations}
  \begin{align}
  \label{eq:varpidef}
  \delta_{\Lambda_{\xi}}\mathbf{\Theta}(\varphi,\delta\varphi)
  & \equiv
    d\varpi_{\xi}\,,
    \\
    & \nonumber \\
    \label{eq:pidef}
    \delta \mathbf{X}(\delta_{\Lambda_{\xi}}\varphi)
    & \equiv
      d\pi_{\xi}\,.
  \end{align}
\end{subequations}

Let us consider $\varpi_{\xi}$ first.  In the case at hands, there are two
kinds of gauge transformations:

\begin{enumerate}

\item Local Lorentz transformations, $\delta_{\sigma}$. The parameter of the
  local Lorentz transformation induced by the diffeomorphism generated by $\xi$
  is

\begin{equation}
  \sigma_{\xi}^{ab}
  =
  \imath_{\xi}\omega^{ab}-P_{\xi}{}^{ab}\,.
\end{equation}

\item Gauge transformations of the 3-form $C$, $\delta_{\chi}$. The
  parameter of the gauge transformation induced by the diffeomorphism
  generated by $\xi$ is

\begin{equation}
  \chi_{\xi}
  =
  \imath_{\xi}C -P_{\xi}\,.
\end{equation}

\end{enumerate}

We find that 

\begin{equation}
    \delta_{\Lambda_{\xi}}\mathbf{\Theta}(\varphi,\delta\varphi)
    \doteq
    d\left[ -\star (e^{a}\wedge e^{b})\delta \sigma_{\xi\, ab}
      +\chi_{\xi}\delta\vartheta\right]
    \equiv
    d\varpi_{\xi}\,.
\end{equation}

As for $\pi_{\xi}$, we find

\begin{equation}
  \delta\mathbf{X}(\delta_{\Lambda_{\xi}}\varphi)
  \doteq
  d\delta\left(- \vartheta P_{\xi}\right)
\equiv
  d\pi_{\xi}
  \,.
\end{equation}

Thus, we find that

\begin{equation}
    \omega(\varphi,\delta\varphi,\delta_{\xi}\varphi)
    \doteq
  d\mathbf{W}[\xi]\,,
\end{equation}

\noindent
with

\begin{equation}
  \mathbf{W}[\xi]
  \equiv
  -\delta\mathbf{Q}[\xi]
  -\imath_{\xi}\mathbf{\Theta}(\varphi,\delta\varphi)
  +\varpi_{\xi}
  -\pi_{\xi}\,,
\end{equation}

\noindent
and we also find that for vector fields $k$ that generate symmetries of all
the fields $\delta_{k}\varphi=0$

\begin{equation}
  d\mathbf{W}[k]
  \doteq
  0\,.
\end{equation}

Integrating this identity over the same hypersurface $\Sigma^{3}$ we
considered for the Smarr formula and using again the Stokes theorem we will
derive the first law, but we must compute $\mathbf{W}[k]$ first. We find

\begin{subequations}
  \begin{align}
    \imath_{k}\mathbf{\Theta}(\varphi,\delta\varphi)
& =    
    -\imath_{k}\star (e^{a}\wedge e^{b})\wedge \delta \omega_{ab}
    -\star (e^{a}\wedge e^{b})\wedge \delta \imath_{k}\omega_{ab}
    +\imath_{k}\star d\phi\delta\phi
    +\imath_{k}C\delta\vartheta\,,
    \\
    & \nonumber \\
    \delta \mathbf{Q}[k]
    & =
    \delta \star (e^{a}\wedge e^{b})P_{k\, ab}
      + \star (e^{a}\wedge e^{b})\delta P_{k\, ab}
       +\delta\left( \vartheta P_{k}\right)\,,    
  \end{align}
\end{subequations}

\noindent
and, therefore, 

\begin{equation}
  \label{eq:W[k]}
  \mathbf{W}[k]
  =
  -\delta \star (e^{a}\wedge e^{b})P_{k\, ab}
  +\imath_{k}\star (e^{a}\wedge e^{b})\wedge \delta \omega_{ab}
  -P_{k}\delta\vartheta
  -\imath_{k}\star d\phi\delta\phi\,,
\end{equation}

\noindent
where we have used Eq.~(\ref{eq:VkvarthetaP}).

As a non-trivial test of all the manipulations we have performed, it can be
checked by an explicit and direct calculation that, indeed, $\mathbf{W}[k]$ is
closed on-shell when the variations of the fields satisfy the linearized
equations of motion and $k$ leaves invariant the background solution. This
calculation can be found in Appendix~\ref{eq:proofdW0}.

Let us proceed to derive the first law:

\begin{equation}
  \begin{aligned}
    0
    & \doteq
    \int_{\Sigma^{3}}d\mathbf{W}[k]
    \\
    & \\
    & =
    \frac{1}{16\pi G_{N}^{(4)}}
    \int_{\Sigma^{3}}d\left\{ -\delta \star (e^{a}\wedge e^{b})P_{k\, ab}
      +\imath_{k}\star (e^{a}\wedge e^{b})\wedge \delta \omega_{ab}
      -\imath_{k}\star d\phi\delta\phi \right\}
    \\
    & \\
    & \hspace{.5cm}
    -\frac{\delta \vartheta}{16\pi G_{N}^{(4)}}\int_{\Sigma^{3}}dP_{k}
    \\
    & \\
    & \doteq
    \frac{1}{16\pi G_{N}^{(4)}}\left\{\int_{S^{2}_{\infty}}
    -\int_{\mathcal{BH}}\right\}\left\{ -\delta \star (e^{a}\wedge e^{b})P_{k\, ab}
      +\imath_{k}\star (e^{a}\wedge e^{b})\wedge \delta \omega_{ab}
      -\imath_{k}\star d\phi\delta\phi \right\}
    \\
    & \\
    & \hspace{.5cm}
    +\frac{\delta \vartheta}{16\pi G_{N}^{(4)}}\int_{\Sigma^{3}}\imath_{k}G
    \\
    & \\
    & \doteq
    \delta M +\tfrac{1}{4}\Sigma \delta\phi_{\infty} -T\delta S
    +\frac{\delta \vartheta}{16\pi G_{N}^{(4)}}\int_{\Sigma^{3}}\imath_{k}\star
    \frac{\partial V}{\partial \vartheta}
    \\
    & \\
    & =
    \delta M +\tfrac{1}{4}\Sigma \delta\phi_{\infty} -T\delta S
    -\Phi_{\vartheta}\delta \vartheta\,,
  \end{aligned}
\end{equation}

\noindent
where we have defined

\begin{equation}
  \Phi_{\vartheta}
  \equiv
  -\frac{1}{16\pi G_{N}^{(4)}}\int_{\Sigma^{3}}\imath_{k}\star
    \frac{\partial V}{\partial \vartheta}\,.  
\end{equation}

Observe that

\begin{equation}
  \Phi_{\vartheta}\delta \vartheta
  =
  \Phi_{\alpha}\delta \alpha\,,  
\end{equation}

\noindent
where $\Phi_{\alpha}$ is the potential that occurs in the Smarr formula
Eq.~(\ref{eq:Smarrformula2}).

In the above derivation we have used Eq.~(\ref{eq:Sigmaversusphi}) and the
vanishing of $k$ over the bifurcation sphere.

Thus, we arrive to our final expression for the first law

\begin{equation}
  \label{eq:firstlaw}
  \delta M
  =
  T\delta S
  -\tfrac{1}{4}\Sigma \delta\phi_{\infty}
  +\Phi_{\alpha}\delta \alpha\,.
\end{equation}

\section{The Anabal\'on-Oliva hairy black hole}
\label{sec-anabalalon-olivaBH}

In this section we study, as an illustration of our general results, the
asymptotically-flat Anabal\'on-Oliva (AO) hairy black hole constructed in
Ref.~\cite{Anabalon:2012ih} as a solution of the system we are considering
Eq.~(\ref{eq:action}) for the particular scalar potential\footnote{Apart from
  the change in the signature of the metric, we have set $\kappa=1/2$ in the
  results of Ref.~\cite{Anabalon:2012ih}.}

\begin{equation}
  \label{VVphi} 
  \begin{aligned}
    V(\phi)
    & =
    \frac{2\alpha}{\nu^{2}} \left\{ \frac{\nu-1}{\nu+2} \sinh{
        \left[\sqrt{\frac{\nu+1}{\nu-1}} \phi\right]} -\frac{\nu+1}{\nu-2}
      \sinh{ \left[\sqrt{\frac{\nu-1}{\nu+1}}\phi \right]}
    \right.
    \\
    & \\
    & \hspace{.5cm}
    \left.
      +4\frac{\nu^{2}-1}{\nu^{2}-4} \sinh{ \left(
          \frac{1}{\sqrt{\nu^{2}-1}}\phi \right)} \right\}\,,
  \end{aligned}
\end{equation}

\noindent
where $\alpha$ is a constant with dimensions of inverse length squared and
$\nu$ is a dimensionless parameter (which means that we are, actually, dealing
with a family of theories) such that $\nu>1$.

The metric and the scalar field of the AO solution are given in
Ref.~\cite{Anabalon:2012ih} in the form

\begin{subequations}
  \begin{align}
    ds^{2}
    & =
      \Omega(r)\left[ F(r) d t^{2}-\frac{d r^{2}}{F(r)}
      -d\Omega^{2}_{(2)}\right]\,,
    \\
    & \nonumber \\
    \phi
    & =
    \sqrt{\nu^{2}-1} \ln{(r/\eta)}\,,
  \end{align}
\end{subequations}

\noindent
where the functions $\Omega$ and $F$ are given by

\begin{subequations}
  \begin{align}
    \Omega(r)
    & =
      \frac{\nu^{2} \eta^{\nu-1}
      r^{\nu-1}}{\left(r^{\nu}-\eta^{\nu}\right)^{2}}\,,
    \\
    & \nonumber \\
    F(r)
    & =
      \frac{r}{\eta\Omega(r)}
      +\alpha\left[\frac{1}{\left(\nu^{2}-4\right)}
      -\left(1+\frac{\eta^{\nu}
      r^{-\nu}}{\nu-2}-\frac{\eta^{-\nu} r^{\nu}}{\nu+2}\right)
      \frac{r^{2}}{\eta^{2} \nu^{2}}\right]\,,
  \end{align}
\end{subequations}

\noindent
but it is convenient to define a new coordinate $\rho$, related to $r$ by

\begin{equation}
  r
  =
  \eta W^{1/\nu}(\rho)\,,
\,\,\,\,\,
\text{where}
\,\,\,\,\,
W(\rho)
\equiv
\left(1-\frac{\nu/\eta}{\rho}\right)\,,
\end{equation}

\noindent
which satisfies

\begin{equation}
  dr
  =
  \frac{d\rho}{\Omega(\rho)}\,.
\end{equation}

\noindent
Then, defining

\begin{equation}
  \lambda
  \equiv
  F\Omega\,,
  \,\,\,\,\,
  \text{and}
  \,\,\,\,\,
  R^{2}
  \equiv
  \Omega\,,
\end{equation}

\noindent
the metric and scalar field of the solution take the form

\begin{subequations}
  \begin{align}
    ds^{2}
    & =
    \lambda(\rho) d t^{2}
    -\lambda^{-1}(\rho)d\rho^{2}
    -R^{2}(\rho) d\Omega^{2}_{(2)}\,,
    \\
    & \nonumber \\
    \phi
    & =
    \frac{\sqrt{\nu^{2}-1}}{\nu} \ln{W}\,,
  \end{align}
\end{subequations}

\noindent
with the functions $\lambda$ and $R$ given by

\begin{subequations}
  \begin{align}
    R^{2}(\rho)
    & =
      W^{1-1/\nu}\rho^{2}\,,
    \\
    & \nonumber \\
    \lambda(\rho)
    & =
      W^{1/\nu}
      +\alpha\left\{\frac{W^{1-1/\nu}}{\left(\nu^{2}-4\right)}
      - \frac{W^{1+1/\nu}}{\nu^{2}}-\frac{W^{1/\nu}}{\nu^{2}(\nu-2)}
      +\frac{W^{2+1/\nu}}{\nu^{2}(\nu+2)}
     \right\}\rho^{2}\,.
  \end{align}
\end{subequations}

In this form the metric is asymptotically flat with the standard normalization
for $\rho\rightarrow \infty$: for large $\rho$

\begin{subequations}
  \begin{align}
    \lambda
    & \sim
      1-\frac{3\eta^{2}+\alpha}{3\eta^{3}\rho}
      +\mathcal{O}(1/\rho^{2})\,,
    \\
    & \nonumber \\
    R^{2}
    & \sim
      \rho^{2} +\mathcal{O}(\rho)\,,
    \\
    & \nonumber \\
    \phi
    & \sim
      -\frac{\sqrt{\nu^{2}-1}}{\eta\rho}+\mathcal{O}(1/\rho^{2})\,,
  \end{align}
\end{subequations}

\noindent
so that the ADM mass $M$, the scalar charge $\Sigma$ and the asymptotic value
of the scalar $\phi_{\infty}$ are given by

\begin{subequations}
  \begin{align}
    M
    & =
      \frac{3\eta^{2}+\alpha}{6\eta^{3}}\,,
    \\
    & \nonumber \\
    \Sigma
    & =
      -\frac{\sqrt{\nu^{2}-1}}{\eta}\,,
    \\
    & \nonumber \\
    \phi_{\infty}
    & =
      0\,.
  \end{align}
\end{subequations}

\noindent
For this asymptotic value of the dilaton, the conditions
Eqs.~(\ref{eq:conditions}) are, indeed, satisfied.

Observe that, since $\alpha$ and $\nu$ are parameters of the theory, the
solution contains only one free parameter, $\eta$, which determines the ADM
mass. The scalar charge $\Sigma$ can, then, be written in terms of the ADM
mass and the parameters that define the theory. Therefore, it describes
secondary hair.

It is not difficult to recover some known metrics: for $|\nu|=1$, the
``hairless limit''\footnote{The potential diverges for $|\nu|=1$ and the
  theory is not well defined for this value of the parameter.} the scalar
vanishes identically and

\begin{subequations}
  \begin{align}
    R^{2}(\rho)
    & =
      \rho^{2}\,,
    \\
    & \nonumber \\
    \lambda(\rho)
    & =
      W
      +\frac{\alpha}{3}\left(W-1\right)^{3}\rho^{2}
      =
      1-\left(\frac{1}{\eta}+\frac{\alpha}{3\eta^{3}}\right)\frac{1}{\rho}\,,
  \end{align}
\end{subequations}

\noindent
which correspond to the Schwarzschild metric with

\begin{equation}
  M
  =
  \frac{3\eta^{2}+\alpha}{6\eta^{3}}\,.
\end{equation}

When $\alpha=0$ the potential vanishes and one recovers the
Janis-Newman-Winicour solutions \cite{Janis:1968zz}

\begin{subequations}
  \begin{align}
    R^{2}(\rho)
    & =
      W^{1-1/\nu}\rho^{2}\,,
    \\
    & \nonumber \\
    \lambda(\rho)
    & =
      W^{1/\nu}\,,
    \\
    & \nonumber \\
    \phi
    & =
    \frac{\sqrt{\nu^{2}-1}}{\nu} \ln{W}\,.
  \end{align}
\end{subequations}

\noindent
The parameter $\nu$ does not occur in the action and it is just an integration
constant related to the mass and scalar charge by

\begin{equation}
  \nu
  =
  \frac{\sqrt{M^{2}+\Sigma^{2}/4}}{M}\,,
\end{equation}

\noindent
and the metric is singular except when the scalar charge vanishes,
($\Sigma=0$,  $|\nu|=1$)

$\lambda(\rho)$ vanishes for $\rho=\nu/\eta$, but so does $R^{2}(\rho)$
(except for $|\nu|=1$, Schwarzschild), which indicates that there is a
singularity there.  It can be shown numerically that, for $\nu<3$ $\lambda$
has another zero at some $\rho_{h}>\nu/\eta$ which converges towards the
singularity at $\rho=\rho_{sing}=\nu/\eta$. Finding an analytical expression
for $\rho_{h}$ in terms of $\alpha,\nu$ and $\eta$ is too complicated, but we
can check the Smarr formula and the results concerning the scalar charge that
we have obtained in the previous sections using the property
$\lambda(\rho_{h})=0$ and $R^{2}(\rho_{h})\neq 0$ only. In the calculations it
is often convenient to use a coordinate\footnote{This coordinate $x$ is
  different from the one defined in Ref.~\cite{Anabalon:2012ih}.}

\begin{equation}
  x \equiv \frac{\nu}{\eta\rho}\,,
\hspace{1cm}
  x_{h} \equiv \frac{\nu}{\eta\rho_{h}}\,.
\end{equation}

Using these properties, we find the surface gravity 

\begin{equation}
  \begin{aligned}
    \kappa
    & =
    \frac{\left[ -\alpha\rho_{h}^{2}+\nu^{2} \left( \nu-2 \right) 
   \right] W^{-1+1/\nu}(\rho_{h})}{2\eta\nu^{2} (\nu-2) \rho_{h}^{2}}
     +\frac{ \alpha\rho_{h} W^{2+1/\nu}(\rho_{h})}{\nu^{2}(\nu+2)}
 \\
 & \\
 & \hspace{.5cm}
 -\frac{\alpha\left[2\eta \left(\nu^{2}-4 \right) \rho_{h}-2\,\nu^{2}+3\,\nu+2 \right]W^{1+1/\nu}(\rho_{h})}{2\eta\,{\nu}^{2} \left( {\nu}^{2}-4 \right)}
 +\frac{\alpha\rho_{h} W^{1-1/\nu}(\rho_{h})}{\left(\nu^{2}-4 \right)}
 \\
 & \\
 & \hspace{.5cm}
    +\frac{\alpha (\nu-1) W^{-1/\nu}(\rho_{h})}{2\eta \left(\nu^{2}-4 \right)}
    -\frac{\alpha\left( 2\,\eta\,\rho_{h}+{\nu}^{2}-
        \nu-2 \right) W^{1/\nu}(\rho_{h})}{2\eta\nu^{2} \left(\nu-2 \right)}\,,  
  \end{aligned}
\end{equation}

\noindent
and the area of the horizon 

\begin{equation}
    A
    =
    4\pi R^{2}(\rho_{h})
    =
    4 \pi W^{1-1/\nu}(\rho_{h})\rho_{h}^{2}\,.
\end{equation}

We can, then, compute

\begin{equation}
  M-2ST
  =
   \frac{3\eta^{2}+\alpha}{6\eta^{3}}-R^{2}(\rho_{h}) \kappa\,, 
\end{equation}

\noindent
which, according to the Smarr formula Eq.~(\ref{eq:Smarrformula2}) should
equal to 

\begin{equation}
\frac{1}{8\pi G_{N}^{(4)}}
\int_{\Sigma^{3}}\imath_{k}\star V
=
-\frac{1}{2G_{N}^{(4)}}
  \int^{\infty}_{\rho_{h}} R^{2}Vd\rho\,,
\end{equation}

\noindent
which is proportional to $\alpha$. We have checked that the Smarr formula
holds in this form.

We have also checked that Eq.~(\ref{eq:esa}) which follows from the
coordinate-independent definition of scalar charge Eq.~(\ref{eq:Sigmadef})
holds.

Finally, checking the first law Eq.~(\ref{eq:firstlaw}) in this solution is
difficult. In order to test the term proportional to the scalar charge one
must have a family of solutions in which the asymptotic value of the scalar is
a free parameter, which is not the case here.

\section{Discussion}
\label{sec-discussion}

In this paper we have derived a Smarr formula and a first law for the extended
thermodynamics of the black-hole solutions of the theories described by the
action Eq.~(\ref{eq:action}) using Wald's formalism and the results of
Refs.~\cite{Elgood:2020svt,Meessen:2022hcg}. Our results coincide with those
of Ref.~\cite{Astefanesei:2023sep} except for the inclusion of the term
proportional to the scalar charge and the variation of the asymptotic value of
the scalar. This term is somewhat mysterious since there are no asymptotically
flat black-hole solutions for asymptotic values of the scalar other than zero,
but it may make sense in a wider class of not asymptotically-flat
solutions. In any case, the term is clearly there since it arises exactly in
the same way as in all the theories considered in
Refs.~\cite{Gibbons:1996af,Ballesteros:2023iqb}.

Using an extension of the covariant definition of scalar charge given in
Refs.~\cite{Pacilio:2018gom,Ballesteros:2023iqb} we have shown that, in the
presence of a bifurcate horizon, the scalar charge is determined by the
parameters of the theory, the particular scalar potential, and the value of
the scalar on the horizon, and should be considered as ``secondary hair''. On
the other hand, a well-defined scalar charge is possible in the presence of a
bifurcate black-hole horizon only if the scalar potential satisfies a set of
quite restrictive conditions given in Eq.~(\ref{eq:conditions}), previously
found in Ref.~ Ref.~\cite{Cadoni:2015gfa} following other considerations.
These conditions are equivalent to a ``no-hair theorem'' for all the theories
whose potentials do not satisfy them. Some scalar profiles such as those
considered in Ref.~\cite{Nucamendi:1995ex} may evade these constraints,
though.

Our results still leave some questions unanswered: are there hairy black-hole
solutions in all the theories whose scalar potentials satisfy all the right
conditions?  What happens in theories with more scalar fields or in theories
in which the scalars couple to curvature scalars? Clearly, much more work is
necessary to find a general pattern of behaviour of scalar fields in black-hoe
spacetimes and some work in this direction is already in progress
\cite{kn:BMOP}.

\section*{Acknowledgments}

TO would like to thank Patrick Meessen and Matteo Zatti for useful
conversations. RB would like to thank Ra\'ul Rojas for his help in the early
stages of this work.  This work has been supported in part by the MCI, AEI,
FEDER (UE) grants PID2021-125700NB-C21 (``Gravity, Supergravity and
Superstrings'' (GRASS)), and IFT Centro de Excelencia Severo Ochoa
CEX2020-001007-S.  The work of RB has also been supported by the National
Agency for Research and Development [ANID] Chile, Doctorado Nacional, under
grant 2021-21211461 and by PUCV, Beca Pasant\'{\i}a de Investigaci\'on.  TO
wishes to thank M.M.~Fern\'andez for her permanent support.

\appendix

\section{Proving that $d\mathbf{W}[k]\doteq 0$}
\label{eq:proofdW0}

In this appendix we give a detailed proof of the on-shell closedness of the
2-form charge given in Eq.~(\ref{eq:W[k]}). We the sequence of steps in which
we use basc geometric properties, equations of motion and symmetry properties
is almost self-descriptive and we will not comment upon them in order not to
extend too much this appendix.

\begin{equation}
  \begin{aligned}
    d\mathbf{W}[k]
    & =
    -\mathcal{D}\delta \left[\star (e^{a}\wedge e^{b})\right] P_{k\, ab}
    -\delta \left[\star (e^{a}\wedge e^{b})\right]\wedge \mathcal{D}P_{k\, ab}
    +\mathcal{D}\left[\imath_{k}\star (e^{a}\wedge e^{b})\right]
    \wedge \delta \omega_{ab}
    \\
    & \\
    & \hspace{.5cm}
    -\imath_{k}\star (e^{a}\wedge e^{b})\wedge  \mathcal{D}\delta  \omega_{ab}
    -dP_{k}\delta\vartheta 
    -d\imath_{k}\star d\phi\delta\phi
    -\imath_{k}\star d\phi\wedge \delta d\phi
    \\
    & \\
    & =
    -\mathcal{D}\delta \left[\star (e^{a}\wedge e^{b})\right] P_{k\, ab}
    +\delta \left[\star (e^{a}\wedge e^{b})\right]\wedge \imath_{k}R_{ab}
    +\mathcal{D}\left[\imath_{k}\star (e^{a}\wedge e^{b})\right]
    \wedge \delta \omega_{ab}
    \\
    & \\
    & \hspace{.5cm}
    -\imath_{k}\star (e^{a}\wedge e^{b})\wedge  \delta  R_{ab}
    +\imath_{k}G\delta\vartheta
    +\imath_{k}d\star d\phi\delta\phi
    -\imath_{k}\star d\phi\wedge \delta d\phi
    \\
    & \\
    & =
    -\delta d\left[\star (e^{a}\wedge e^{b})\right] P_{k\, ab}
    +2\omega^{a}{}_{c}\delta \left[\star (e^{c}\wedge e^{b})\right] P_{k\, ab}
    \\
    & \\
    & \hspace{.5cm}
    +\delta \left[\star (e^{a}\wedge e^{b})\right]\wedge \imath_{k}R_{ab}
    -2P_{k}{}^{a}{}_{c}\star (e^{c}\wedge e^{b})
    \wedge \delta \omega_{ab}
    \\
    & \\
    & \hspace{.5cm}
    -\imath_{k}\star (e^{a}\wedge e^{b})\wedge  \delta  R_{ab}
    +\imath_{k}\star \frac{\partial V}{\partial\vartheta}\delta\vartheta
    +\imath_{k}\star \frac{\partial V}{\partial\phi} \delta\phi
    -\imath_{k}\star d\phi\wedge \delta d\phi
  \end{aligned}
\end{equation}

\begin{equation}
  \begin{aligned}
    d\mathbf{W}[k]
    & =
    -2\delta \left[\omega^{a}{}_{c}\wedge \star (e^{c}\wedge e^{b})\right]
    P_{k\, ab}
    +2\omega^{a}{}_{c}\delta \left[\star (e^{c}\wedge e^{b})\right] P_{k\, ab}
    \\
    & \\
    & \hspace{.5cm}
    +\delta \left[\star (e^{a}\wedge e^{b})\right]\wedge \imath_{k}R_{ab}
    -2P_{k}{}^{a}{}_{c}\star (e^{c}\wedge e^{b})
    \wedge \delta \omega_{ab}
    \\
    & \\
    & \hspace{.5cm}
    -\imath_{k}\star (e^{a}\wedge e^{b})\wedge  \delta  R_{ab}
    +\imath_{k}\star \frac{\partial V}{\partial\vartheta}\delta\vartheta
    +\imath_{k}\star \frac{\partial V}{\partial\phi} \delta\phi
    -\imath_{k}\star d\phi\wedge \delta d\phi
    \\
    & \\
    & =
-\delta\left\{ \imath_{k}\star (e^{a}\wedge e^{b})\wedge  R_{ab}\right\}
    +\imath_{k}\left[\delta \star (e^{a}\wedge e^{b})\wedge   R_{ab}\right]
    \\
    & \\
    & \hspace{.5cm}
    +\imath_{k}\star \frac{\partial V}{\partial\vartheta}\delta\vartheta
    +\imath_{k}\star \frac{\partial V}{\partial\phi} \delta\phi
    -\imath_{k}\star d\phi\wedge \delta d\phi
  \end{aligned}
\end{equation}

\begin{equation}
  \begin{aligned}
    d\mathbf{W}[k]
    & =
    -\delta\left\{ -\tfrac{1}{2} \imath_{k}\star d\phi\wedge d\phi
      +\imath_{k}\star V\right\}
    +\imath_{k}\left[\delta \star (e^{a}\wedge e^{b})\wedge   R_{ab}\right]
    \\
    & \\
    & \hspace{.5cm}
    +\imath_{k}\star \frac{\partial V}{\partial\vartheta}\delta\vartheta
    +\imath_{k}\star \frac{\partial V}{\partial\phi} \delta\phi
    -\imath_{k}\star d\phi\wedge \delta d\phi
    \\
    & \\
    & =
    \tfrac{1}{2} \imath_{k}\delta\star d\phi\wedge d\phi
    +\tfrac{1}{2} \imath_{k}\star d\phi\wedge \delta d\phi
     -\imath_{k}\delta \star V
    +\imath_{k}\left[-\imath_{c}\star (e^{a}\wedge e^{b})\wedge   R_{ab}\wedge
    \delta e^{c}\right]
    \\
    & \\
    & \hspace{.5cm}
    +\imath_{k}\star \frac{\partial V}{\partial\vartheta}\delta\vartheta
    +\imath_{k}\star \frac{\partial V}{\partial\phi} \delta\phi
    -\imath_{k}\star d\phi\wedge \delta d\phi\,,
  \end{aligned}
\end{equation}

\begin{equation}
  \begin{aligned}
    d\mathbf{W}[k]
    & =
    \tfrac{1}{2} \imath_{k}\delta\star d\phi\wedge d\phi
    +\tfrac{1}{2} \imath_{k}\star d\phi\wedge \delta d\phi
     -\imath_{k}\delta \star V
    \\
    & \\
    & \hspace{.5cm}
    +\imath_{k}\left\{\left[\tfrac{1}{2} \imath_{a}\star d\phi\wedge d\phi
        +\tfrac{1}{2}\star d\phi \imath_{a}d\phi
        -\imath_{a}\star V
\right]\wedge
    \delta e^{a}\right\}
    \\
    & \\
    & \hspace{.5cm}
    +\imath_{k}\star \frac{\partial V}{\partial\vartheta}\delta\vartheta
    +\imath_{k}\star \frac{\partial V}{\partial\phi} \delta\phi
    -\imath_{k}\star d\phi\wedge \delta d\phi
    \\
    & \\
    & =
    \tfrac{1}{2} \imath_{k}\delta\star d\phi\wedge d\phi
    -\tfrac{1}{2} \imath_{k}\star d\phi\wedge \delta d\phi
     -\imath_{k}\delta \star V
    \\
    & \\
    & \hspace{.5cm}
    +\imath_{k}\left\{\left[\tfrac{1}{2} \imath_{a}\star d\phi\wedge d\phi
        +\tfrac{1}{2}\star d\phi \imath_{a}d\phi
\right]\wedge \delta e^{a}\right\}
    \\
    & \\
    & \hspace{.5cm}
    \imath_{k}\left\{
      -\imath_{c}\star V\wedge \delta e^{c}
      +\star \frac{\partial V}{\partial\vartheta}\delta\vartheta
          +\imath_{k}\star \frac{\partial V}{\partial\phi} \delta\phi
\right\}
    \\
    & \\
    & =
    \tfrac{1}{2} \imath_{k}\delta\star d\phi\wedge d\phi
    -\tfrac{1}{2} \imath_{k}\star d\phi\wedge \delta d\phi
     -\imath_{k}\delta \star V
    \\
    & \\
    & \hspace{.5cm}
    +\imath_{k}\left\{\left[\tfrac{1}{2} \imath_{a}\star d\phi\wedge d\phi
        +\tfrac{1}{2}\star d\phi \imath_{a}d\phi
\right]\wedge
    \delta e^{a}\right\}
    \\
    & \\
    & \hspace{.5cm}
    +\imath_{k}\delta \star V
    \\
    & \\
    & =
    \tfrac{1}{2} \imath_{k}\left[\left(-\imath_{a}\star d\phi\wedge d\phi
        -\star d\phi\imath_{a}d\phi\right)\wedge \delta e^{a} +\star d\delta
      \phi\wedge d\phi\right]
    -\tfrac{1}{2} \imath_{k}\star d\phi\wedge \delta d\phi
    \\
    & \\
    & \hspace{.5cm}
    +\imath_{k}\left\{\left[\tfrac{1}{2} \imath_{a}\star d\phi\wedge d\phi
        +\tfrac{1}{2}\star d\phi \imath_{a}d\phi
\right]\wedge
    \delta e^{a}\right\}
    \\
    & \\
    & =
0\,.
  \end{aligned}
\end{equation}


\end{document}